\documentclass{aa}
\usepackage{psfig}

\begin{document}

\title{The very local Hubble flow: computer simulations of dynamical history}

\author{
Arthur D. Chernin\inst{1,2,3}
\and
Igor D. Karachentsev\inst{4}
\and
Mauri J. Valtonen\inst{2,5}
\and
Valentin P. Dolgachev\inst{1}
\and
Ludmila M. Domozhilova\inst{1}
\and
Dmitry I. Makarov\inst{4,6}
}

\institute{ 
Sternberg Astronomical Institute, Moscow University, Moscow, 119899, Russia
\and 
Tuorla Observatory, Turku University,Piikki\"o, 21 500, Finland
\and 
Astronomy Division, University of Oulu, 90014, Finland
\and 
Special Astrophysical Observatory, Nizhnii Arkhys, 369167, Russia
\and 
University of West Indies, Trinidad and Tobago
\and  
Isaac Newton Institute of Chile, SAO Branch, Russia
}

\authorrunning{A.D. Chernin et al.}
\titlerunning{The very local Hubble flow}

\date{Received / Accepted}

\abstract{ The phenomenon of the very local ($\le3$~Mpc) Hubble
flow is studied on the basis of the data of recent precision
observations. A set of computer simulations is performed to trace
the trajectories of the flow galaxies back in time to the epoch
of the formation of the Local Group. It is found that the `initial
conditions' of the flow are drastically different from the linear
velocity-distance relation. The simulations enable also to
recognize the major trends of the flow evolution and identify the
dynamical role of universal antigravity produced by cosmic vacuum.
 }

\maketitle

\section{Introduction}

As is well-known, the original Hubble diagram plots galaxy
kinematics for the distances within 20~Mpc, after the correction
of a systematic error in the determination of distances. Sandage
(1999) confirms that the Hubble flow takes its origin at very
small distances, 1.5--2~Mpc, from the center of the Local Group
(see also Ekholm et al. 2001). A recent high precision mapping of
the very local velocity field has covered the spatial scales
between 1.5--2 and 3~Mpc (Karachentsev et al. 2000, 2002, 2003).
High precision has become possible due to remarkable progress in
accurate distance measurements for galaxies in the vicinity of
the Local Group (LG), --- mostly due to observations with the
{\em Hubble Space Telescope}. The velocity field has been found
by Karachentsev and co-workers to have a fairy regular
kinematical structure with the linear velocity-distance relation
and the expansion rate of $72\pm15$~km/s/Mpc. The flow is rather
cold: its one-dimensional mean random motion is about 30~km/s.
The expansion flow on these spatial scales is referred to as the
very local Hubble flow (hereafter VLHF).

In this paper, we use the recent precision data (Karachentsev et
al. 2002) to follow the VLHF dynamical history. We have performed a
set of computer simulations for the present, past and also future
of VLHF. This enable us to re-construct the `initial conditions'
for VLHF  at the epoch of the Local Group formation 12.5~Gyr ago
and found that the observed fairly regular state of the flow is a
result of the dynamical evolution from a highly disordered and
violent initial state. We found that the initial state of VLHF
resembles a model of the Little Bang proposed by Byrd, Valtonen,
McCall and Innanen (1994) for the early Local Group. This state
is in general agreement as well with a new picture of the Local
Group formation discussed recently by van den Bergh (2003); this
picture involves also violent dynamics as a key physical factor
of the process.

In Sec.2, a theory background is discussed which takes into account the
dynamical effect of newly discovered cosmic vacuum; in Sec.3, the basic
data we use are summarized; the simulations are presented and analyzed
in Sec.4; conclusions are given in Sec.5.

\section{The Hubble-Sandage paradox}

The phenomenon of VLHF has not been predicted by the cosmological
theory. Moreover, the existence of the cosmological expansion in
the local volume contradicts widely accepted cosmological
concepts. Indeed, it has commonly been believed that the very
notion of the cosmological expansion is applicable to only very
large spatial distances, and so only when one reaches the scale
of great clusters of galaxies (100--300 Mpc) one should find the
markers that participate in the cosmological expansion. An
obvious reason for this is that the expansion with the linear
velocity-distance relation is directly associated with the
uniformity of the universe. And the matter distribution is
uniform on the spatial scales larger than 100--300 Mpc.
Observations reveal no uniformity in the nearby spatial
distribution of galaxies, in the scale range from a few to
20~Mpc. Meanwhile, the cosmological expansion was originally
discovered deep inside the cell of uniformity in the galaxy
distribution.

A question arises: how the observed spatial non-uniformity of the galaxy
distribution in the local volume may be compatible with the observed
regular linear velocity field?

Sandage (1986; see also Sandage et al. 1972) was the first who
discussed such a controversy, and, according to his recent
conclusion, an ``explanation of why the local expansion field is so
noiseless remains a mystery'' (Sandage 1999). It is also puzzling
that the local rate of expansion is similar to the global one, if
not exactly the same, within 10--15 percent accuracy (Sandage 1999).

The linear velocity-distance relation in local and global
expansion flows and the almost (if not exactly) the same
expansion rate (the Hubble parameter) in the both indicate that
there is a common physical agent that affects the expansion flow
from the distances of a few Mpc up to the observation horizon. It
has been proposed (Baryshev, Chernin, Teerikorpi 2001; Chernin
2001, Chernin, Teerikorpi, Baryshev 2002; Chernin, Karachentsev,
Teerikorpi 2003) that this physical agent is cosmic vacuum (or the
cosmological constant, or dark energy) with its perfectly uniform
energy density on all spatial scales. We have argued that this
idea offers a possible solution to the Hubble-Sandage paradox
that has existed in cosmology for more than 70 years.

Cosmic vacuum has been discovered in recent SN Ia observations
(Riess et al. 1998; Perlmutter et al. 1999) confirmed by all the
bulk of cosmological evidence (see for a fresh review Peebles and
Ratra 2003). The vacuum density $\rho_V$ comprises up to 70-75
percent the total density of the Universe. The dynamical effect
of cosmic vacuum is enhanced by the fact that, according to the
Friedmann theory, the effective gravitating (actually,
antigravitating) density of vacuum is $\rho_V + 3p_V = -2
\rho_V$, where $p_V = - \rho_V$ is the vacuum pressure.

Our suggestion above is invoked by Thim, Tammann, Saha, Dolphin,
Sandage, Tolstoy and Labhard (2003) in a recent treatment of their
new observations on the extreme quietness of the local (1--10
Mpc) expansion field. They also mention that the suggestion makes a
continual precision mapping of the local velocity even more
crucial.

\section{Basic data on VLHF}

In the dataset on the very local velocity field published by
Karachentsev et al. (2002), there are 38 galaxies located within
3 Mpc. Two of them have no measured velocities yet; two other do
not have estimated distances; six more have only low accuracy
distances. Six of the rest with distances around 2.8 Mpc are
located on the front side of the Canes Venatici cloud and
apparently move from us toward the cloud center with an
additional velocity of about 85~km~s$^{-1}$. With their
exclusion, the collection of
 22 galaxies (including one located in the center of the Canes Venatici cloud)
is accepted as the observational basis for the computer
simulations. These galaxies may reasonably be considered as `most
typical representatives' for the very local Hubble flow (VLHF).
Their names, distances and velocities relative to the center of
the Local Group are given in Table 1 (columns 2--4). Note
that the galaxy distances, $R$, are known with typical accuracy of
10\%. The galaxies are small in mass (dwarfs) and fairly
separated from each other, as is seen from both distances and
position angles; because of this, their interaction with each
other is negligibly weak compared to the interaction with the two
major galaxies of the Local Group (hereafter LG) and vacuum (see
below). The 22 galaxies reveal together the Hubble
velocity-distance linear relation, $V = H_L R$, with the time
rate $H_L = 72 \pm 15$~km~s$^{-1}$~Mpc$^{-1}$ and the
one-dimensional velocity dispersion $30$~km~s$^{-1}$. For more
detailed data on the VLHF galaxies and their analysis (including
a discussion of the accuracy of the observations) see in
Karachentsev et al. (2002).

\begin{table}
\caption{Galaxies of the very local Hubble flow}
\centerline{
\begin{tabular}{rlrrrr}
\hline
   N  &Name     & R      & V    & $R_0$  & $V_0$  \\
      &         &Mpc     &km/s  &Mpc     &km/s\\
\hline
   1  & SagDIG  &  1.15&    23 &     0.49&   140\\
   2  & SexB    &  1.63&   111 &     0.60&   162\\
   3  & Antlia  &  1.70&    66 &     1.23&   188\\
   4  & N3109   &  1.70&   110 &     1.16&   167\\
   5  & SexA    &  1.74&    94 &     0.69&   177\\
   6  & KKR25   &  1.79&    68 &     0.83&    95\\
   7  & E294-010&  1.92&    81 &     0.82&   104\\
   8  & KKH98   &  2.02&   151 &     0.97&   144\\
   9  & KK230   &  2.03&   126 &     0.31&   172\\
  10  & N300    &  2.11&   114 &     0.62&   140\\
  11  & UA438   &  2.16&    99 &     0.92&   109\\
  12  & I5152   &  2.18&    75 &     1.27&    77\\
  13  & GR8     &  2.37&   136 &     0.60&   178\\
  14  & U8508   &  2.55&   186 &     0.15&   221\\
  15  & I3104   &  2.62&   171 &     0.49&   214\\
  16  & N404    &  2.63&   195 &     0.71&   186\\
  17  & DD0187  &  2.69&   172 &     0.56&   190\\
  18  & DD0190  &  2.83&   263 &     0.54&   269\\
  19  & KKH86   &  2.92&   209 &     0.33&   231\\
  20  & GamB    &  3.00&   266 &     0.45&   268\\
  21  & N1560   &  3.05&   171 &     1.08&   156\\
  22  & N2403   &  3.09&   268 &     0.31&   277\\
\hline
\end{tabular}
}
\end{table}

\section{Computer simulations}

\subsection{The model}

We have developed a set of computer simulations in which
individual galaxies of VLHF do not interact with each other; the
observational reason for this is seen from what was said in the
section above. We also take into account that the mass of LG
(including dark mass) is strongly concentrated (Karachentsev et
al. 2002) to the two major galaxies of the group. We assume that
dark matter halo of the Milky Way (MW) and the Andromeda Galaxy
(AG) are spherical. The adopted total mass of MW is
$1\times10^{12} M_{\sun}$ and the total mass of AG is
$1.5\times10^{12} M_{\sun}$. In this statement of the problem, a
galaxy of VLHF is considered moving in the gravity field of the
two major galaxies of LG (including their dark matter haloes) and
the antigravity field of cosmic vacuum. Therefore the computer
simulations are reduced to the integration of the Newtonian
restricted tree-body problem on the cosmic vacuum background, for
each of the 22 galaxies of the Table 1. The galaxy velocities are
assumed to be radial, at the present state of the flow.

Cosmic vacuum is represented in the simulations by a `medium'
with a perfectly uniform energy density which is also constant in
time, as it follows from the Friedmann model. The concordance
figure (see again for a review Peebles and Ratra, 2003) for the
vacuum energy density is $\rho_V = (0.7\pm0.1) \rho_c$, where
$\rho_c = 2 \times 10^{-29} h^2$~g~cm$^{-3}$ is the critical
density estimated with the `global' Hubble constant $h =
H/100$~km~s$^{-1}$~Mpc$^{-1}$; $h = 0.65 \pm 0.10$. The present
cosmic age is assumed to be 14 Gyr.

The simple Kahn-Woltjer model (Kahn and Woltjer 1959) for LG
which assumes the straight linear relative motion of the two
major galaxies of the group is re-computed with the account of
the new data on the galaxy masses and the vacuum density. The
present separation 0.7 Mpc and the relative velocity
$-120$~km~s$^{-1}$ are adopted. The two major galaxies, MW and AG,
started their motion toward each other 12.5 Gyr ago. In our
computer simulations, the trajectories of the VLHF galaxies are
traced back in time to that moment in the past. The trajectories
are also computed for about 6 Gyr in the future, up to the moment
when MW and AG come into contact collision.

\begin{figure*}
\centerline{\psfig{figure=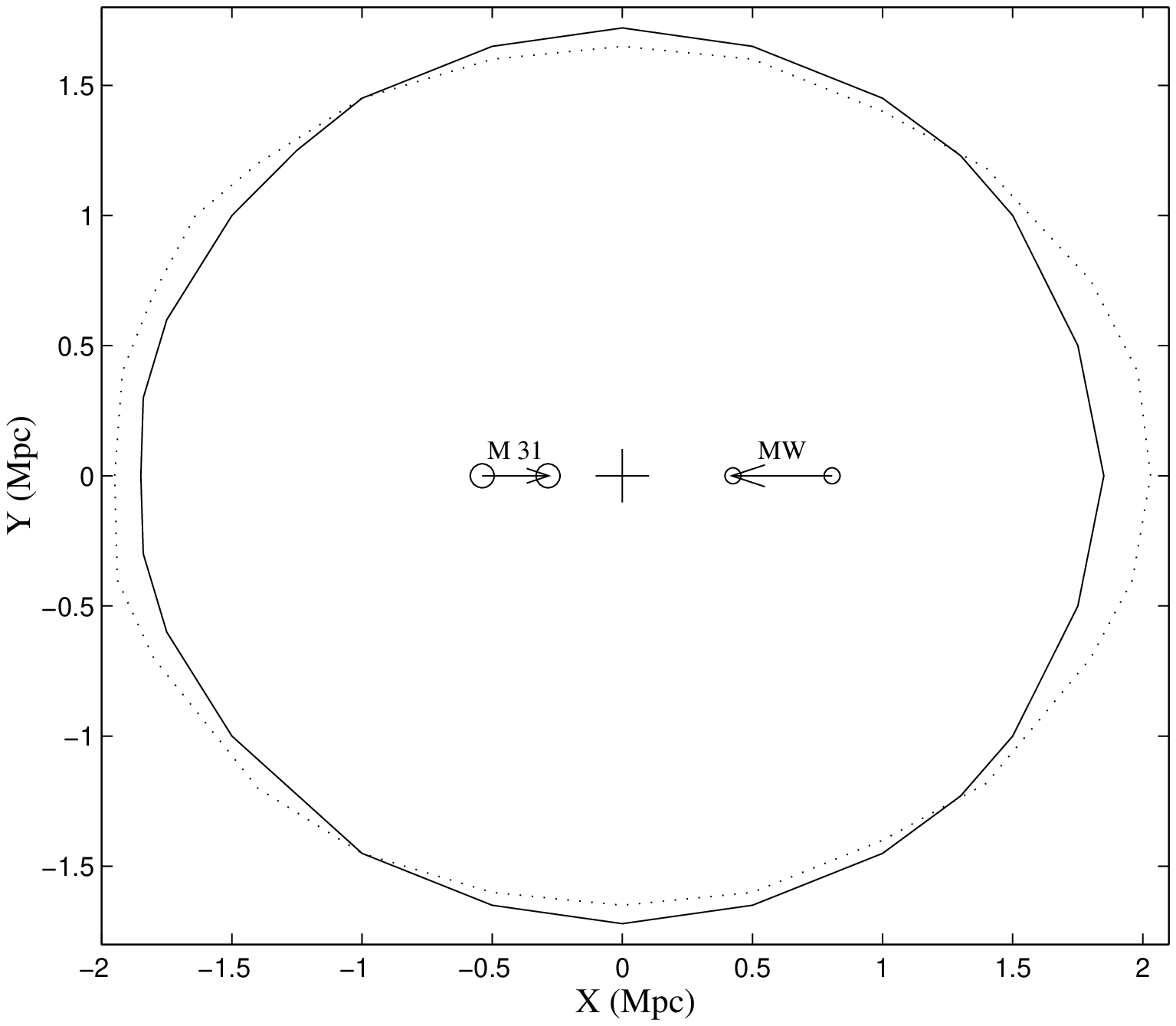,width=\textwidth}}
\caption{Zero-gravity surface around the Local Group now (solid
line) and 12.5 Gyr ago (dashed line).\label{f:zg}}
\end{figure*}

Antigravity of cosmic vacuum dominates dynamically (it is taken
into account that its effective energy density is $-2\rho_V$ ---
see Sec.2) during all the 12.5 Gyr history of the Local Group at
the distances larger than 2 Mpc from the center of the group.
This is one of the results  of the computer three-body problem.
The critical `zero-gravity surface', i.e. the surface at which
the radial component of the gravity and antigravity forces are
exactly balanced, is showed in Fig.~\ref{f:zg}. It may be seen
from the figure that the surface can be embedded completely
between two concentric (centered to the center of the Local
Group) spheres, one with the radius of 1.8 Mpc and the other 1.7
Mpc, at present. The two enveloping spheres have radii 2 Mpc and
1.6 Mpc in the past, 12.5 Gyr ago. Thus, the surface is nearly
spherical, and it remains nearly unchanged during all the history
of VLHF. Outside the zero-gravity sphere the potential is
repulsive, and it can be considered as nearly spherically
symmetrical and nearly static, with a  good accuracy. This
dynamical background determines the major features of the VLHF
evolution, in our simulations.

\begin{figure*}
\centerline{\psfig{figure=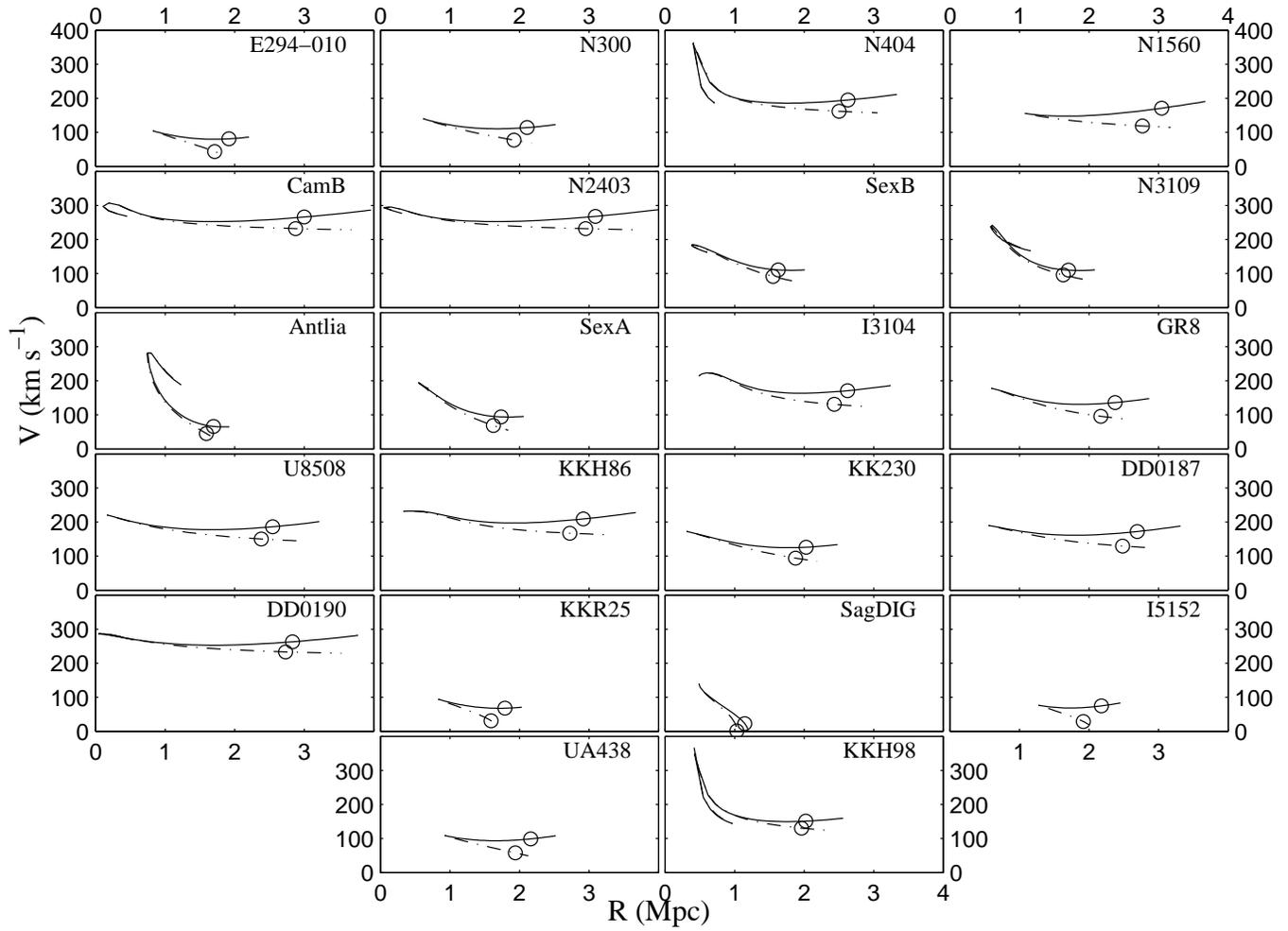,width=\textwidth}}
\caption{The phase portrait of the the very local Hubble flow:
radial velocity vs. radial distance plots for 22 galaxies of the
flow -- solid lines.  For comparison: same for trajectories
calculated with the same initial conditions 12.5 Gyr ago, but
without cosmic vacuum -- dashed lines. Circles indicate the
present state.\label{f:phases}}
\end{figure*}

The results of the simulations are presented in Table 1 and
Figs.~2--4. In Table 1 (columns 5,6), the initial state of VLHF
is described by the radial velocities and distances of the flow
galaxies 12.5 Gyr ago. The dynamics of VLHF is illustrated by the
VLHF phase portrait (Fig.~\ref{f:phases}) which consists of 22
evolutionary curves in the velocity-distance plot; these are the
radial velocities and radial distances relative to the center of
LG. The dynamical role of cosmic vacuum may be recognized from
the comparison of the real phase curves with `imaginary' phase
curves that are computed for the same initial conditions 12.5 Gyr
ago, but with no vacuum background. As one may see, a typical
phase curve describes a decrease  of the velocity with the growth
of the distance from the LG center, at the first stage of the
evolution. At the next stage, the velocity grows with distance
under the action of the cosmic vacuum antigravity.

\subsection{VLHF initial conditions: the Little Bang}

 The initial state of VLHF as is recognized from the simulations is
 drastically different from any naive expectations
that could treat VLHF as a primeval cosmological flow that might
be only slightly distorted initially. The radial velocities and
radial distances of the galaxies at the initial moment 12.5 Gyr
ago given in columns 5,6 of Table 1 suggest a conclusion that the
initial state of VLHF has nothing in common with such a picture.
The deviations from an imaginary `unperturbed' initial flow that
could exist on the same spatial scales at the same early time are
very strong.

To examine how robust this conclusion may be, we have performed a
special set of test simulations at which an additional transverse
velocity is assumed that is 20--30 percent of the observed radial
velocity of the galaxies at present. We also have repeated
simulations with variations of the observed distances, $R$, within
$\pm10\%$. The result has demonstrated a good qualitative
agreement with the basic conclusion about a highly perturbed
initial state of the flow 12.5 Gyr ago, in both cases. The
structure of the initial states in such test simulations differs
from that of the basic simulation only in quantitative details.

The most striking fact is that
a substantial fraction of the trajectories, 9 of 22,
take start on the `other side' of the MW-AG line of centers. The galaxies
with these initial conditions move toward the center of the Local Group,
initially, so that their initial velocities are negative, in Table 1.
Their flow is a flow of contraction, not expansion, at that time.
The galaxies with negative initial velocities gain considerable infall
velocities near the center of the Local Group, that reach 180--300~km~s$^{-1}$.
Then they pass the central region of the group and begin
to move from the center (continuing their motion in the same direction in
space). In this way, the initial contraction flow transforms into the
expansion flow.

During this transformation, galaxies gain also an additional
velocity, --- now it is a positive velocity of recession. The
acceleration of this nature is not due to antigravity of vacuum;
this is exactly the same dynamical effect of gravitational
acceleration that was studied in details for an early dynamics of
LG in the model of the Little Bang (Byrd et al. 1994). Violent
gravitational interactions of the VLHF galaxies with MW and AG and
LG as a whole are also similar to another picture of the early LG
described recently by van den Bergh (2003).

Together with 7 galaxies that move outward the center initially
with high (around 200~km~s$^{-1}$) velocities (most probably,
they were also accelerated earlier in the same manner), these 9
galaxies form a fast sub-flow of VLHF. A slow sub-flow of 6
galaxies starts its expansion with the velocities of
80--160~km~s$^{-1}$. The initial conditions for both fast and
slow sub-flows occupy an area in the radial velocity space from
$-277$ to $+231$~km~s$^{-1}$ and an area in the radial distance
space from 0.2 to 1 Mpc. Therefore, the initial spread of the
velocities is measured by a figure of 650~km~s$^{-1}$ at that
time. For comparison, at present the same 22 galaxies have
velocities within a much narrower interval from 23 to
268~km~s$^{-1}$, and so the spread is measured by 245~km~s$^{-1}$.

There are no signs of the linear regularity in the
velocity-distance relation, in the initial state of VLHF. For
instance, the expansion rate (the ratio $\dot R/R$) estimated for
individual trajectories prove to be in the interval from $-932$
to 700 ~km~s$^{-1}$~Mpc$^{-1}$, initially. So the initial spread
of this quantity is measured by 1600~km~s$^{-1}$~Mpc$^{-1}$(!).
This is a clear quantitative measure of highly disordered
nonlinear initial structure of the flow. It may be compared with
the present-day state of VLHF, at which the expansion rate is
observed from 22 to 87~km~s$^{-1}$~Mpc$^{-1}$, and so the spread
is only 65~km~s$^{-1}$~Mpc$^{-1}$.

The phase portrait (Fig.~\ref{f:phases}) of the initial VLHF
reveals complex dynamics that can be understood within the
framework of the Little Bang (Byrd et al. 1994). According to
this framework, the formation of the Local Group and nearby
galaxies, including ones that constitute now VLHF, is due to
violent dynamics involving close passings, contact collisions and
merging of many sub-galactic units in the volume of 1--2 Mpc
across. In this process, the major fraction of the material falls
into two major potential wells formed by the dark matter
concentrations in the volume, while the VLHF galaxies represent
only debris that AG and probably MW as well ejected from their
common potential well into outer volume. The physical mechanisms
of ejection are studied in details by Byrd et al. (1994).

It seems most probable that only accelerated ejected fragments
(dwarf galaxies and sub-galactic units) were able to survive as
individual physical objects in this violent environment and
escape from the LG potential well out of the zero-gravity sphere.
If so, the Little Bang dynamics was mainly responsible for the
origin of the VLHF galaxies and for the initial conditions of
their motions. The quantitative results given by Byrd et al.
(1994) indicate that the velocities and distances in the initial
state of VLHF (see columns 5,6 of Table 1) are quite feasible for
the ejected bodies, in the violent dynamics of the Little Bang.

Note that some concrete features of the original version of the
Little Bang model need to be re-considered now in the light of new
observational data. However the idea of the violent dynamics for
the early LG is in quite good agreement with the current data. A
recently published picture (van den Bergh 2003) for the LG origin
demonstrates the naturalness of the initial violent dynamics in a
clear way.

\subsection{Evolution of the Hubble  ratio}

\begin{figure*}
\centerline{\psfig{figure=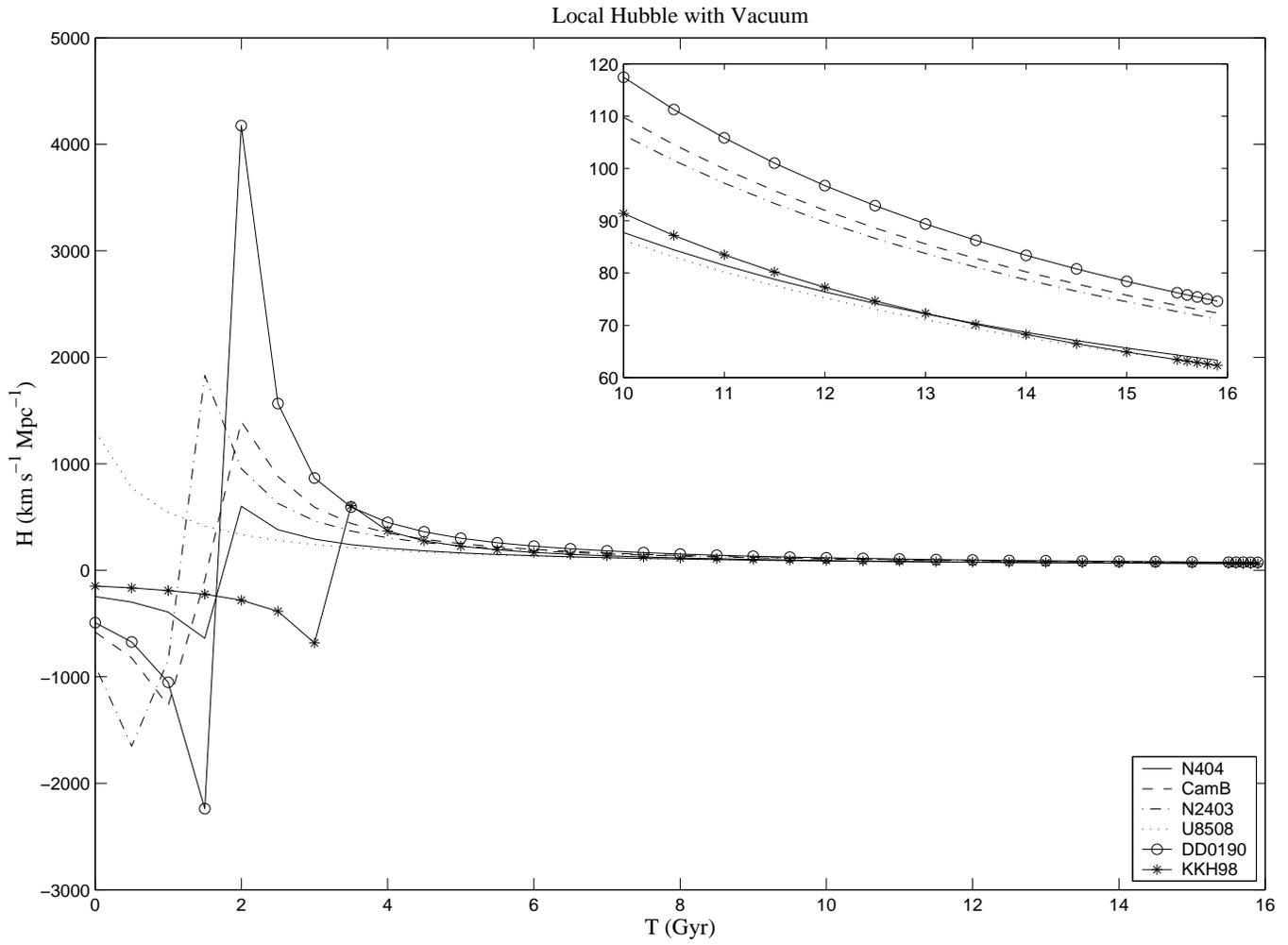,width=\textwidth,angle=-90}}
\caption{Velocity-distance ratio for the galaxies of the very
local Hubble flow as a function of time.\label{f:hvacuum}}
\end{figure*}

Fig.~\ref{f:hvacuum} shows the Hubble ratio, or the time rate,
$H_L = V/R$, which is the individual velocity-distant ratio for
the VLHF galaxies, as functions of time. The convergence of the
bunch of the 22 trajectories to the universal time rate is
obvious from the figure: this is the major trend of the VLHF
evolution which makes VLHF be essentially a cosmological
phenomenon.

\begin{figure*}
\centerline{\psfig{figure=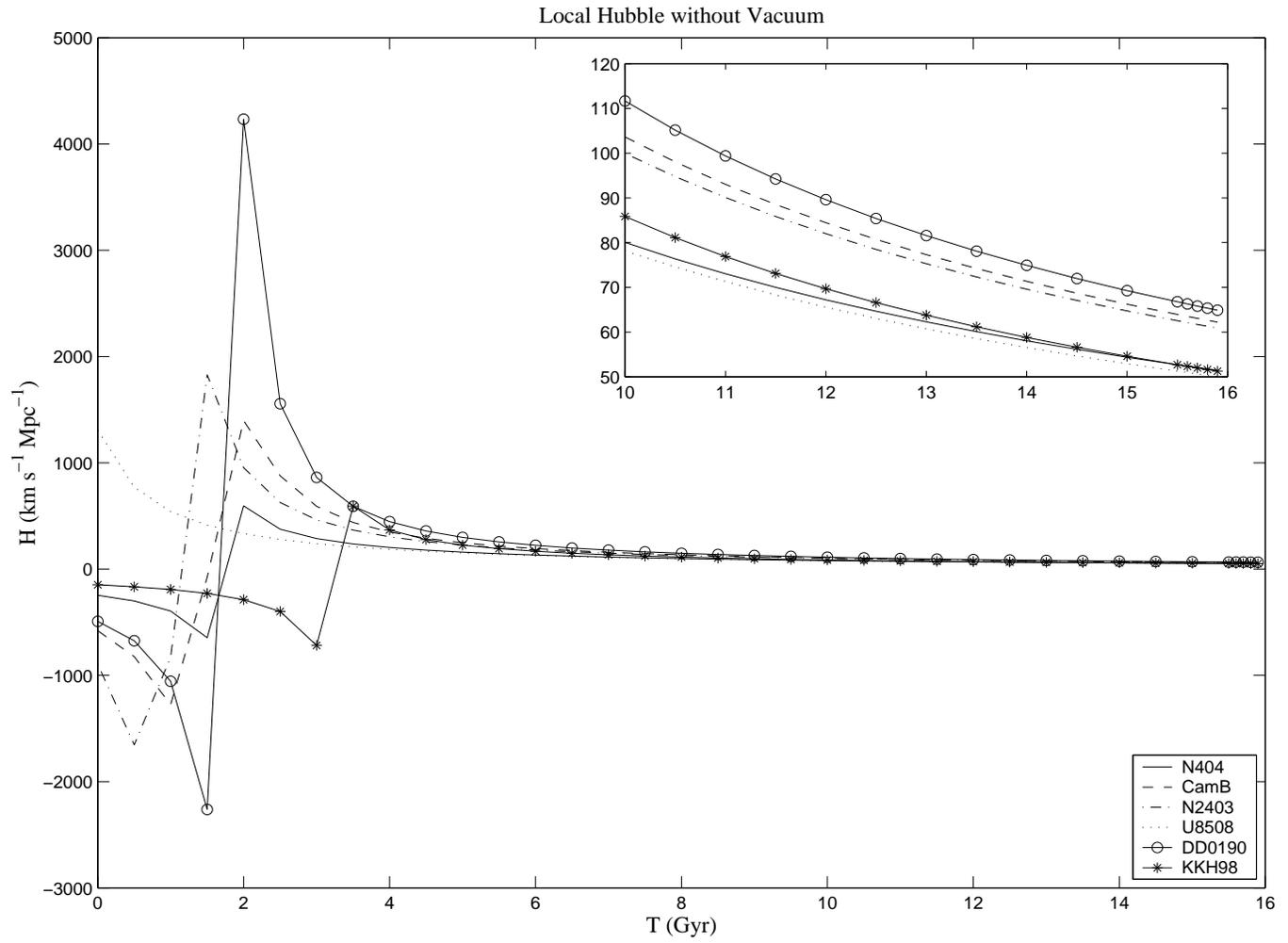,width=\textwidth,angle=-90}}
\caption{For comparison: same for trajectories calculated with the
same initial conditions 12.5 Gyr ago, but without cosmic
vacuum.\label{f:hclassic}}
\end{figure*}

Fig.~\ref{f:hclassic} shows the same for the `imaginary' (no
vacuum) trajectories. As is seen from the plots, the role of
vacuum increases systematically with time, while the role of LG
gravity is only decreasing. Meanwhile the same trend as in
Fig.3 reveals in Fig.4 as well: even in the model without vacuum,
the individual expansion rates tend to converge to a common one
for all the galaxy sample. (The difference is only in numbers, and
the dynamical effect of vacuum gives a higher mean Hubble ratio
at present.) The similarity is completely due to the initial
conditions which are the same for both models and -- which is
seemingly more important -- resulted in the backwards calculations
from the rather smooth observed Hubble law. Indeed, there is,
generally, no reason to expect that a regular linear flow would
arise from arbitrary initial distributions of distances and
velocities for bodies moving in the gravitational field of the
Local Group without vacuum.

In an additional set of simulations, we tried an example of a
`random' initial velocity-distance distribution for bodies in the
close vicinity of the LG. Some of the bodies were captured by the
LG gravity, while the others escaped and moved away from the
group. It was found that the flow of the escaped bodies revealed
an evolution to the regular linear velocity-distance relation
only in the presence of cosmic vacuum. This trend was most
obvious for larger ($> 3$ Mpc) distances from the center of the
LG.

In the presence of vacuum, two effects are especially important.
First, vacuum accelerates the motions of the VLHF galaxies by its
antigravity. Second, acting as a time independent dynamical
factor, vacuum tends to supply each individual galaxy of VLHF
with one and the same expansion rate, independently of the galaxy
initial kinematical states. Both physical effects are obvious
from the consideration of the asymptotical state of the flow, in
the limit of large times, when the role of gravity vanishes and
antigravity controls the flow completely.

In this limit, the solution for any individual radial trajectory
has a form: $R(t) \propto \exp(H_V t); t \rightarrow \infty$,
where $H_V = (\frac{8\pi G}{3} \rho_V)^{1/2}$. Therefore each
individual time rate $H(t)_L$ tends with time to $V/R = \dot R/R
= H_V = Const$ for any trajectory. Both effects mentioned above
lead finally to the formation of the flow with the linear
velocity-distance law:  $\dot R = H_V R$, where the common time
rate for the VLHF galaxies is constant in space and time and
determined by vacuum only: $H_L = H_V$.

As for the Hubble global flow, similar considerations show that
its asymptotical expansion rate is also determined by vacuum
only: $H(t)$ tends to $ H_V$ when $t \rightarrow \infty $. Thus,
asymptotically, VLHF and the global expansion flow become
completely identical in their kinematical structure. This is the
net ultimate result of the universal antigravity of cosmic vacuum
on all spatial scales.

It seems especially remarkable that the present states of both
VLHF and the global flow are not very far from the asymptotical
state; this is seen, first of all,  from the fact that the
present-day observational value of $H_L = 72 \pm
15$~km~s$^{-1}$~Mpc$^{-1}$ and the present-day observational
value for $H = 65 \pm 10$  are both fairly close to the
theoretical limit $ H_V = 55 \pm 10$~km~s$^{-1}$~Mpc$^{-1}$
(actually, all the three values are compatible with a common
figure near, say, 60~km~s$^{-1}$~Mpc$^{-1}$).

\section{Conclusions}

Until quite recently, the structure and dynamics of the galaxy
flow around the Local Group have remained poorly known because of
the lack of reliable data on distances to most of the nearby
galaxies. The recent high accuracy measurements of these
distances have led to the discovery of the real structure of the
fairly regular very local ($ \le 3$ Mpc) Hubble flow (Karachentsev
et al. 2000, 2002, 2003). Basing on these data, we have started
herein detailed quantitative studies of the physical nature of the
phenomenon. An approach we try is suggested by the recent
discovery of cosmic vacuum (Riess et al. 1998, Perlmutter et al.
1999). We have argued earlier (see the references in Sec.2) that
cosmic vacuum is a key dynamical factor not only in the Universe
as a whole, but also in our close vicinity in space where VLHF is
observed.

As a first step in concrete realization of this approach, we have
performed computer simulations of the history of the flow, its
present and future states. The results of the simulations and
their analysis have revealed two basic aspects of the dynamics of
VLHF:

A) The force field that controls VLHF during almost all its
history is dominated by the antigravity of cosmic vacuum at
distances 1.5--2 Mpc from the Local Group center of mass. The
ultimate dynamical state of the flow is entirely determined by
cosmic vacuum with its perfect uniformity. The perfectly regular
antigravity force field introduces regularity to the flow. The
dynamical effect of cosmic vacuum leads asymptotically to the
universal and constant in time expansion rate $H_V = (\frac{8\pi
G}{3}\rho_V)^{1/2} = 55 \pm 10 $~km~s$^{-1}$~Mpc$^{-1}$. The
present state of the flow is not far from its asymptotical state
because its observed Hubble rate is near the asymptotical value
$H_V$.

B) The evolutionary history of VLHF starts at the epoch of the
Local Group formation some 12.5 Gyr ago. At that time, the flow
galaxies, together with the forming major galaxies of the group
and many sub-galactic units, participated in violent nonlinear
dynamics with collisions and merging. VLHF was formed by
relatively small units that survive accretion by the major
galaxies and managed to escape from the gravitational potential
well of the Local Group. Our simulations show that a typical VLHF
member galaxy gained escape velocity from the highly
non-stationary gravity fields of the forming group and a velocity
larger than some 200~km~s$^{-1}$ enabled it to reach the
vacuum-dominated outer region. The simulations we produced do not
describe the violent dynamics of the forming Local Group. However
they give definite indications to the very existence of this
dynamics. It is a special complex problem to re-construct the
violent initial dynamics in the local volume in all its
completeness; the Little Bang model (Byrd et al. 1994) and the
picture presented by van den Bergh (2003) provide important
insights to the problem and give the basic grounds for such a
study.

The approach developed in this paper can be extended (and we will
report the results later) to larger volumes around the Local
Group. One can expect both a similarity to VLHF and some specific
differences for the distances, say 10--100 Mpc which are still
within the cell of uniformity of the galaxy spatial distribution.
The observed bulk motion with 500--600~km~s$^{-1}$ velocity is
one of the major features on these scales. The differences may be
mostly in the initial conditions for the flow on these scales.
But the similarity may definitely be due to cosmic vacuum with its
universal antigravity. It is perfectly uniform cosmic vacuum that
is suggested to be the major physical agent affecting the
expansion flow everywhere (including the bulk motion --- Chernin,
2001), from a few Mpc to the observation horizon.

Another interesting direction for further computational
studies is provided by an opportunity of a more general form of
cosmic antigravity which is due to dark energy with a time variable
density. It was argued in Baryshev et al. (2001) that a variable
dark energy, especially such `coupled' with matter would better
explain the small local velocity than the classical vacuum; this
was because of the fact that the gravity dominated region was
then smaller in the past. In this case, the model for VLHF would
include a decreasing dark energy density which would make the flow
dynamical background essentially non-stationary, -- contrary to
the model presented above.

 The authors are grateful to Yury Efremov and Pekka Teerikorpi
for critical comments and productive suggestions.


\end{document}